\documentclass[10pt,conference]{IEEEtran}
\title{Detecting Argument-Swap Bugs Using Context-Enhanced Code Representations}
\author{
\IEEEauthorblockN{
Subrata Das$^{*}$, Ali Aman$^{\dagger}$, Muhammad Asaduzzaman$^{\dagger}$, Kawser Wazed Nafi$^{\ddagger}$, Salimur Choudhury$^{\S}$
}
\IEEEauthorblockA{
$^{*}$\textit{Department of Computer Science, Lakehead University}; sdas18@lakeheadu.ca
}
\IEEEauthorblockA{
$^{\dagger}$\textit{School of Computer Science, University of Windsor}; \{burkia, masaduzz\}@uwindsor.ca
}
\IEEEauthorblockA{
$^{\ddagger}$\textit{Department of Computer Engineering, Polytechnique de Montreal}; kawser.wazed-nafi@polymtl.ca
}
\IEEEauthorblockA{
$^{\S}$\textit{School of Computing, Queen's University}; s.choudhury@queensu.ca
}
}
\usepackage{soul}
\usepackage{amssymb}
\usepackage{amsmath}
\usepackage{multirow}
\usepackage{listings}
\usepackage{xcolor}
\usepackage{booktabs}
\usepackage{tikz}
\usepackage{url}
\usepackage[hidelinks]{hyperref}
\newcommand{\fdot}{\tikz\fill (0,0) circle (2pt);}
\lstdefinestyle{pythonstyle}{
    language=Python,
    basicstyle=\ttfamily\small,
    keywordstyle=\color{blue},
    stringstyle=\color{magenta!80},
    commentstyle=\color{gray},
    showstringspaces=false,
    columns=flexible,
    numbers=left,
    numberstyle=\small\color{gray},
    stepnumber=1,
    numbersep=6pt,
    frame=single,
    framesep=4pt,
    rulecolor=\color{black!20},
    captionpos=b,
    xleftmargin=5pt,
    xrightmargin=5pt,
    literate={~}{{\textasciitilde}}1,
}

\begin{document}
\maketitle
\begin{abstract}
Names of source code elements convey rich semantic information and have been widely used in software engineering tasks such as bug detection, code completion, type prediction, and code classification. Prior studies exploit lexical similarity between method arguments and formal parameter names to detect bugs caused by incorrectly ordered arguments, typically relying on establishing mappings between method calls and their corresponding definitions. However, such mappings are often difficult to obtain in dynamically typed languages like Python.
In this paper, we present \textbf{BugProbe}, a learning-based approach for detecting incorrectly ordered arguments in Python method calls that does not require call--definition mappings. Our approach leverages multiple sources of contextual information, including local context and argument usage context, and combines name-based similarity with machine learning to construct expressive representations of method arguments. We collect a new dataset of 132,739 Python source files from the top-1,000 starred GitHub repositories, yielding 3,371,244 synthetic training examples, and contribute a curated benchmark of 55 real-world argument-swap bugs manually verified from commit histories. We evaluate our approach on this dataset and show that it achieves high accuracy and consistently outperforms a state-of-the-art baseline across standard evaluation metrics. These results demonstrate that effective detection of argument-ordering bugs is possible without relying on explicit call--definition resolution, making the approach well suited for dynamically typed language settings.
\end{abstract}
\section{Introduction}
\label{sec:intro}

Argument-swap bugs are a common source of programming errors, particularly in dynamically typed languages. Such bugs occur when a developer accidentally passes arguments to a method in the wrong order. We use the term \emph{call site} to refer to a location in source code where a function or method is invoked. Because swapped arguments often have compatible types, the program may continue executing without raising an error, silently producing incorrect behavior that is difficult to diagnose.

Figure~\ref{fig:argswap_bug} illustrates an argument-swap bug where \texttt{resize\_image} expects width followed by height, but the call site mistakenly reverses them. Since both parameters share the same type, this error is type-correct yet semantically invalid, making detection challenging even in statically typed languages. In dynamically typed languages, where static type information is typically unavailable, such bugs are particularly hard to identify.

Identifier names convey valuable semantic information about program behavior. Meaningful names aid program comprehension and support tasks such as code completion \cite{li2017code, liu2020multi, raychev2014code} and identifier name suggestion \cite{lin2017investigating, wainakh2021idbench}. Prior work has also observed that developers often use similar names for call site arguments and their corresponding formal parameters~\cite{liu2016nomen}. As a result, the similarity between argument names and parameter names has been exploited to detect argument-swap bugs~\cite{pradel2013name, liu2016nomen, scott2020out, patra2022nalin}.

\begin{figure}[h]
\centering
\begin{lstlisting}[language=Python]
# Intended usage
def resize_image(width, height):
    return Image.resize(width, height)

# Buggy call: arguments accidentally swapped
new_img = resize_image(img_height, img_width)
\end{lstlisting}
\caption{An argument-swap bug where swapped arguments have compatible types but incorrect semantics}
\label{fig:argswap_bug}
\end{figure}

However, approaches based solely on lexical similarity are fragile. Developers may use semantically related but lexically different identifiers (e.g., \texttt{area} and \texttt{region}), limiting the effectiveness of string-based comparisons. To address this, Pradel et al. proposed DeepBugs, which learns vector representations (embeddings) of identifiers to detect semantic similarity even when names differ lexically \cite{pradel2018deepbugs}.

Despite its effectiveness, DeepBugs has important limitations when applied to dynamically typed languages such as Python. First, it requires mapping call sites to their corresponding method definitions to compare argument names with formal parameter names. This mapping is challenging in dynamic languages and often infeasible when definitions reside in different files or external libraries. In our study of Python code, we found that 89\% of call sites have their definitions located in different files, frequently within third-party libraries. Second, DeepBugs learns embeddings from a limited local token window around every identifier, regardless of its role. Argument-swap detection requires understanding how arguments at call sites relate to one another across broader code structures, which narrow identifier-level contexts may miss.

In this paper, we propose \textbf{BugProbe}, a novel machine learning framework for detecting argument-swap bugs that overcomes these limitations. We formulate detection as binary classification and learn identifier embeddings without requiring call-to-definition mapping. Instead of relying on local token windows, our approach constructs embeddings specifically from argument contexts at call sites, capturing long-term dependencies and structural usage patterns across the codebase.

We evaluate our approach on a large collection of Python code, replicating DeepBugs for Python as a baseline. Our results show that BugProbe effectively detects argument-swap bugs without relying on method definitions at all, achieving an $F_1$ score approximately 11.3\% higher (relative) than DeepBugs across all call sites.

This paper makes the following contributions:
\begin{itemize}
    \item A call-site-centered analysis formulation that constructs argument representations directly from call site contexts, avoiding the need to map call sites to their definitions --- a mapping step that is particularly challenging in dynamically typed languages.
    \item A novel embedding approach that captures long-term dependencies and structural usage contexts of identifier names.
    \item A dataset of 132,739 Python files from the top-1,000 starred GitHub repositories, comprising 3,371,244 synthetic examples across 1,685,622 call sites.
    \item A curated benchmark of 55 real-world argument-swap bugs manually verified from commit histories across 36 open-source repositories.
    \item An empirical evaluation demonstrating significant improvements over DeepBugs on both synthetic and real-world benchmarks.
\end{itemize}

The remainder of this paper is organized as follows. Section~\ref{sec:related} discusses related work. Section~\ref{sec:data} describes data collection. Section~\ref{sec:method} presents the proposed framework. Section~\ref{sec:eval} reports evaluation results. Section~\ref{sec:threats} discusses threats to validity, and Section~\ref{sec:conclusion} concludes the paper.
\section{Related Work}
\label{sec:related}

This section reviews prior research related to detecting argument-ordering and argument-selection defects, as well as studies on argument recommendation and type inference that are relevant to our work.

\textbf{Detecting incorrectly ordered arguments.}
A number of studies have investigated the problem of identifying incorrectly ordered arguments and related defects. Pradel and Gross~\cite{pradel2013name} introduced an anomaly-based approach for detecting swapped arguments of identical types. Their method analyzes recurring call site patterns and flags calls whose argument order deviates from commonly observed naming conventions in similar contexts. Liu et al.~\cite{liu2016nomen} explored the lexical similarity between argument names and formal parameter names in 60 real-world Java programs, demonstrating that name similarity can be leveraged to identify incorrectly ordered arguments and recommend corrections. Li et al.~\cite{li2018lexical} further analyzed lexical similarity distributions, name length, and causes of dissimilarity, and discussed applications such as argument renaming and argument suggestion.

Rice et al.~\cite{rice2017detecting} proposed a name-based similarity matching approach to identify argument selection defects, showing that discrepancies between argument and parameter names often indicate potential bugs. Scott et al.~\cite{scott2020out} introduced SWAPD, which combines coverage-based and statistical checking strategies to detect swapped arguments. While effective in specific scenarios, these approaches primarily rely on handcrafted heuristics or lexical similarity measures and do not employ learning-based models to capture richer contextual information.

The most closely related work to ours is DeepBugs, which formulates bug detection as a binary classification problem using learned embeddings derived from local token context. However, DeepBugs requires mapping call sites to their corresponding definitions, which is challenging in dynamically typed languages such as Python. Moreover, it relies on a narrow local context window when constructing embeddings. In contrast, our approach does not depend on call--definition mappings and incorporates multiple complementary usage contexts to construct richer representations of arguments and call sites.

\textbf{Argument recommendation and completion.}
Our work is also related to research on argument recommendation and method call completion. Zhang et al.~\cite{zhang2012automatic} proposed \emph{Precise}, which collects argument usage examples from existing code and matches call-site contexts against a usage database to recommend appropriate arguments. Asaduzzaman et al.~\cite{asaduzzaman2015parc} introduced PARC, a parameter recommendation system that supports a broader range of parameter expression types than Precise. More recently, Nguyen et al.~\cite{nguyen2023arist} combined program analysis with statistical language models to recommend type-compatible arguments.

Although these systems focus on assisting developers during code completion, their recommendations can implicitly reveal argument selection or ordering issues. For example, discrepancies between recommended and actual arguments may signal potential defects. Unlike these approaches, our work explicitly targets argument-ordering bugs and evaluates detection performance directly.

\textbf{Type inference for dynamically typed languages.}
Finally, our study is related to research on type inference for Python programs, which can provide valuable information for detecting argument-related defects. Allamanis et al.~\cite{typilus_2022} proposed Typilus, which uses graph neural networks to model dependencies between code tokens and infer types. Pradel et al.~\cite{typewriter_2020} introduced TypeWriter, which combines probabilistic type prediction with type validation. Mir et al.~\cite{type4py_2022} presented Type4Py, a hierarchical neural model that outperforms both Typilus and TypeWriter. Peng et al.~\cite{10.1145/3510003.3510038} proposed HiTyper, which integrates static analysis with learning-based type recommendations. More recently, Wei et al.~\cite{wei2023typet5} introduced TypeT5, a transformer-based model for type inference, and Peng et al.~\cite{generative_type_inference_2024} proposed TypeGen, which leverages large language models and chain-of-thought prompting for generative type inference.

The inferred types produced by these approaches can be beneficial when detecting incorrectly ordered arguments, particularly when arguments share the same static type. However, our proposed method does not rely on any external type inference system. Instead, it focuses on leveraging contextual usage information directly extracted from source code, making it applicable even when reliable type information is unavailable.

\section{Data Collection}
\label{sec:data}

We collected a dataset of Python source files by cloning the top-1,000 Python repositories from GitHub, ranked by star count. To ensure repository quality and activity, we applied the following inclusion criteria: at least two contributors, at least 100 commits, and at least one commit within the past year. We further retained only files that parse successfully under Python~3.11, ensuring compatibility with our analysis pipeline. This initial collection yielded 381,244 Python source files.

We then applied a series of preprocessing steps to remove files unsuitable for training. First, files containing no source lines or no parseable statements were discarded, removing 12,807 files. Next, files containing no qualifying call sites (i.e., no call sites with at least two arguments) were excluded, removing a further 48,613 files. Code duplication has been shown to negatively affect the performance of machine learning models trained on source code~\cite{Allamanis10Dup}. Following prior work~\cite{mt4py2021}, we therefore deduplicated the remaining 319,824 files using CD4Py~\cite{mt4py2021}\footnote{\url{https://github.com/saltudelft/CD4Py}}, a code deduplication tool for Python. CD4Py tokenizes each file, represents it using TF--IDF vectors, and clusters near-duplicate files using a $k$-nearest-neighbor search; a single representative file was retained per cluster, removing 17\% of files (54,370 files). We applied size-based filtering (kept files between 2 KB and 25 KB) to keep training computationally feasible, which removed a further 132,715 files.

After preprocessing, the final dataset consists of 132,739 Python source files drawn from the 1,000 repositories, comprising 34,357,043 total source lines. The dataset contains 1,685,622 qualifying call sites. Each call site contributes one positive (correct) example and one negative (incorrect) example generated by swapping the first two arguments, yielding 3,371,244 total examples. Following the 5-fold cross-validation split described in Section~\ref{sec:eval}, 2,696,686 examples are used for training and 674,558 for testing in each fold. Table~\ref{tab:dataset-stats} summarizes the key statistics of the final dataset.

\begin{table}[htb]
\centering
\caption{Dataset Statistics}
\label{tab:dataset-stats}
\renewcommand{\arraystretch}{0.9}
\setlength{\tabcolsep}{6pt}

\begin{tabular}{l r l r}
\toprule
\textbf{Statistic} & \textbf{Value} & \textbf{Statistic} & \textbf{Value} \\
\midrule
Repositories            & 1,000     & Total examples        & 3,371,244 \\
Source files            & 132,739   & Training examples     & 2,696,686 \\
Total source lines      & 34,357,043& Test examples         & 674,558 \\
Qualifying call sites   & 1,685,622 & Real-world bugs       & 55 \\
\bottomrule
\end{tabular}
\end{table}

\textbf{Call-Site Argument Arity: }Our negative example generation swaps only the first two positional arguments at a call site, which raises the question of how much of the qualifying call-site population this covers. Among the 1,685,622 qualifying call sites (i.e., each call site contains two or more arguments), 1,308,770 (77.6\%) have exactly two positional arguments and are therefore fully covered by our (1,2)-swap negative sampling strategy. The remaining 376,852 call sites (22.4\%) have three or more positional arguments, meaning that at least one possible argument-position pair at these call sites is not exercised by our synthetic negative examples, and for which BugProbe's generalization to higher-arity swaps remains untested. We do not further analyze whether classification difficulty correlates with call-site arity or argument position (e.g., parameters with default values, which tend to appear at the end of the parameter list), due to space limitations; we leave this to future work. We view extending negative example generation to arbitrary argument-position pairs, and evaluating whether representations learned from (1,2)-swaps transfer to higher-arity swaps, as an important future direction.

We additionally note that keyword arguments\footnote{\url{https://docs.python.org/3/tutorial/controlflow.html\#keyword-arguments}} are excluded from our pipeline by design rather than as a limitation: a keyword argument names its parameter explicitly at the call site (e.g., \texttt{f(height=h, width=w)}), so reordering keyword arguments does not change the binding and cannot produce an argument-swap bug. Of the 7,143,411 total call sites identified across our corpus, 646,819 (9.1\%) pass all arguments by keyword and are therefore structurally immune to the bug class we target; these are correctly excluded from our qualifying call sites rather than mishandled. Variadic call sites (\texttt{*args}/\texttt{**kwargs}) are harder to resolve statically and are therefore treated as ordinary positional calls in our pipeline, a simplification we note as a limitation.

\textbf{Real-World Bug Dataset: }In addition to the synthetic dataset, we contribute a curated dataset of 55 real-world argument-swap bugs collected from the commit histories of the 1,000 repositories. For each repository, we mined commits that modified function call sites and identified candidate changes where the ordering of arguments had been altered. To ensure that each candidate reflected a genuine bug fix rather than a refactoring, we explicitly verified that the same commit did not also modify the corresponding function definition in a way that reordered its parameter declarations --- such a co-change would indicate an intentional renaming or signature refactor rather than a correction of a misplaced argument. Remaining candidates were then manually verified by two human annotators, who examined the commit message for explicit mention of an argument-ordering mistake and inspected the surrounding code context to confirm that the original ordering constituted a defect.

The 55 bugs span 36 repositories, which are a subset of the 1,000 repositories in our collection. To prevent data leakage, all source files containing one of these real-world bugs are excluded from both the training and test splits used in our synthetic evaluation, and consequently from Word2Vec training as well. The real-world bug dataset is used exclusively as a held-out benchmark to assess the practical detection ability of BugProbe, as described in Section~\ref{sec:eval}.
\section{Methodology}
\label{sec:method}

In this section, we explain how we reproduce DeepBugs in Python, and how our proposed framework, BugProbe, works and addresses the limitations of DeepBugs.

\subsection{DeepBugs Reproduction}

DeepBugs was originally designed for JavaScript. Its core idea—learning name-based bug detectors using identifier embeddings and supervised classification—is language-agnostic. We therefore adapt the approach to Python while preserving the original problem formulation and learning pipeline. 

\subsubsection{Data Extraction}

For each Python source file in the corpus, we parse the Abstract Syntax Tree (AST) and traverse it to identify calls with at least two arguments. For every such call site, we extract a tuple of name-based features analogous to those used in DeepBugs. The extracted tuple includes:
\begin{itemize}
    \item The name $n_{\textit{base}}$ of the base object if the call is a method call, or an empty string otherwise.
    \item The name $n_{\textit{callee}}$ of the called function.
    \item The names $n_{\textit{arg}1}$ and $n_{\textit{arg}2}$ of the first and second argument.
    \item The types $t_{\textit{arg}1}$ and $t_{\textit{arg}2}$ of the first and second argument for arguments that are literals, or empty strings otherwise.
    \item The names $n_{\textit{param}1}$ and $n_{\textit{param}2}$ of the formal parameters of the called function, or empty strings if unavailable.
\end{itemize}

Each extracted tuple constitutes a positive example, which is assumed to be correct under the assumption that most code in the corpus is correct. To generate negative examples without manual labeling, we swap the first and second argument names and types. This transformation is likely to introduce an argument-ordering bug while preserving the surrounding context. The resulting dataset therefore consists of balanced pairs of positive (correct) and negative (incorrect) examples. The exact tuple structures used to construct the training examples are summarized in Table~\ref{tab:tuple_patterns}.

\begin{table*}[t]
\centering
\caption{Tuple patterns used for generating correct and incorrect training examples.}
\label{tab:tuple_patterns}
\normalsize
\setlength{\tabcolsep}{6pt}
\begin{tabular}{l l p{12.5cm}}
\toprule
\textbf{Framework} & \textbf{Label} & \textbf{Tuple structure} \\
\midrule
DeepBugs & Correct &
$\langle
\mathit{n}_{\mathit{base}},
\mathit{n}_{\mathit{callee}},
\mathit{n}_{\mathit{arg}1},
\mathit{n}_{\mathit{arg}2},
\mathit{t}_{\mathit{arg}1},
\mathit{t}_{\mathit{arg}2},
\mathit{n}_{\mathit{param}1},
\mathit{n}_{\mathit{param}2}
\rangle$ \\
DeepBugs & Incorrect &
$\langle
\mathit{n}_{\mathit{base}},
\mathit{n}_{\mathit{callee}},
\mathbf{\mathit{n}_{\mathit{arg}2}},
\mathbf{\mathit{n}_{\mathit{arg}1}},
\mathbf{\mathit{t}_{\mathit{arg}2}},
\mathbf{\mathit{t}_{\mathit{arg}1}},
\mathit{n}_{\mathit{param}1},
\mathit{n}_{\mathit{param}2}
\rangle$ \\
\midrule
BugProbe & Correct &
$\langle
\mathit{LC}_{\mathit{callee}},
\mathit{LC}_{\mathit{arg}1},
\mathit{AUC}_{\mathit{arg}1},
\mathit{PCLU}_{\mathit{arg}1},
\mathit{LC}_{\mathit{arg}2},
\mathit{AUC}_{\mathit{arg}2},
\mathit{PCLU}_{\mathit{arg}2},
\mathit{t}_{\mathit{arg}1},
\mathit{t}_{\mathit{arg}2}
\rangle$ \\
BugProbe & Incorrect &
$\langle
\mathit{LC}_{\mathit{callee}},
\mathbf{\mathit{LC}_{\mathit{arg}2}},
\mathbf{\mathit{AUC}_{\mathit{arg}2}},
\mathbf{\mathit{PCLU}_{\mathit{arg}2}},
\mathbf{\mathit{LC}_{\mathit{arg}1}},
\mathbf{\mathit{AUC}_{\mathit{arg}1}},
\mathbf{\mathit{PCLU}_{\mathit{arg}1}},
\mathbf{\mathit{t}_{\mathit{arg}2}},
\mathbf{\mathit{t}_{\mathit{arg}1}}
\rangle$ \\
\bottomrule
\end{tabular}
\end{table*}

\subsubsection{Embedding Generation}

To enable semantic reasoning over identifier names, we learn distributed vector representations of identifiers and literals using an unsupervised Word2Vec model, following the DeepBugs approach. Python source files are traversed via their ASTs and converted into linear token sequences containing only identifiers and literals, with all other syntactic elements discarded. Each file is treated as a single sentence, and identifiers and literals are disambiguated using distinct prefixes.

We train the embeddings using the Continuous Bag-of-Words (CBOW) architecture with a context window of size 20 and an embedding dimensionality of 200. We also limit the vocabulary to 10{,}000 most frequent tokens, with all others mapped to a shared \textit{unknown} token. The resulting embeddings capture semantic similarities between names and are learned independently of the downstream classification task.

\subsubsection{Classifier}

Each positive and negative example is mapped to a fixed-length vector by concatenating the vector representations of its constituent elements. Identifier-based fields are represented using their learned 200-dimensional embeddings, while literal type information is encoded using fixed-length binary vectors. This concatenation yields a uniform vector representation for all call sites.

These vectors are used to train a supervised binary classifier implemented as a feedforward neural network. Following DeepBugs, the network consists of a single hidden layer of size 200 and a sigmoid output layer, with dropout applied to reduce overfitting. The model is trained using binary cross-entropy loss and RMSprop optimization, and is evaluated on a held-out test set to detect likely argument-ordering bugs in unseen Python code.

\subsection{Proposed Framework}

\begin{figure*}[t]
    \centering
    \includegraphics[width=1\textwidth]{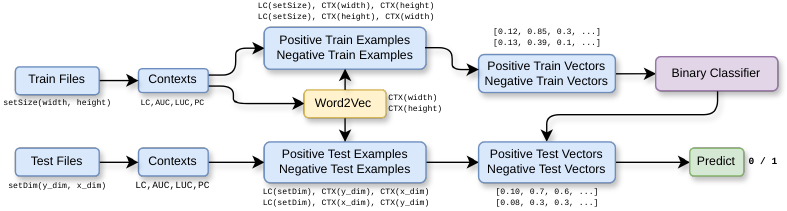}
    \caption{Overview of BugProbe.}
    \label{fig:bugprobe_overview}
    \vspace{-2mm}
\end{figure*}

We propose \textit{BugProbe}, a framework that follows the same high-level paradigm as prior work—namely, the combination of unsupervised representation learning with supervised classification. An overview of the BugProbe framework is shown in Figure~\ref{fig:bugprobe_overview}.

Motivated by preliminary experiments showing that broader usage and structural signals improve over name-only features (quantified per-source in RQ3, Section~\ref{sec:eval}), BugProbe substantially modifies the context used during embedding learning. Specifically, our framework is designed to capture long-range contextual information rather than being limited to local syntactic contexts. This allows the learned embeddings to encode more global program semantics, which is critical for accurately identifying subtle bug patterns.

In addition, we redesign the structure of the generated training examples to better align with the embedding learning process. The tuple patterns used in BugProbe are tailored to reflect the broader contextual scope of the embeddings, resulting in more informative and consistent representations. Furthermore, BugProbe requires no explicit call-to-definition mapping, making it robust to incomplete or unavailable static analysis.

Collectively, these design choices enable BugProbe to achieve substantially stronger and more accurate performance compared to DeepBugs across our evaluation benchmarks.

\subsubsection{Data Extraction}

For each Python source file in the corpus, we parse the Abstract Syntax Tree (AST) and traverse it to identify calls with at least two arguments. For every such call site, we extract a tuple of context-based features. The extracted tuple includes:
\begin{itemize}
    \item The local context $LC_{{\textit{callee}}}$ of the call site.
    \item The argument usage context $AUC_{{\textit{arg}1}}$ and $AUC_{{\textit{arg}2}}$ of the first and second argument.
    \item The parent and latest usage context $PCLU_{{\textit{arg}1}}$ and $PCLU_{{\textit{arg}2}}$ of the first and second argument.
    \item The types $t_{{\textit{arg}1}}$ and $t_{{\textit{arg}2}}$ of the first and second argument for arguments that are literals, or empty strings otherwise.
\end{itemize}

Each extracted tuple constitutes a positive example, which is assumed to be correct under the assumption that most code in the corpus is correct. To generate negative examples without manual labeling, we swap the first and second argument contexts and types. This transformation is likely to introduce an argument-ordering bug while preserving the surrounding context. The resulting dataset therefore consists of balanced pairs of positive (correct) and negative (incorrect) examples. These four context levels are described as follows.

\textbf{Local Context (LC)} is a token window around the occurrence (argument or call site). We extract up to 10 tokens before and 10 tokens after the occurrence from the tokenized source line(s), filtering out common punctuation and operators.

\textbf{Argument Usage Context (AUC)} captures how the argument name is used earlier within the same enclosing scope (e.g., a function body). We collect token sequences from preceding lines that reference the argument name, capturing assignments, updates, and prior calls.

\textbf{Latest Usage Context (LUC)} consists of tokens from the closest preceding line in the same scope that mentions the argument name. This approximates the most recent usage signal, such as the last assignment or transformation before the call.

\textbf{Parent Context (PC)} represents the syntactic container of the call site. We identify a meaningful AST parent (e.g., loop header, conditional header, or function definition) and extract tokens from the parent’s line, capturing structural cues that influence argument semantics.

Since PC is derived from the call site itself rather than from a specific argument, it is identical for both arguments at the same call site and would provide no discriminative signal if used alone. We therefore combine PC with LUC into a single per-argument field, $PCLU$, by concatenating their token sequences. This gives each argument a unique representation that captures both the structural context of the call site and the argument’s most recent usage.

\begin{figure}[t]
    \centering
    \hspace*{-1cm}
    \includegraphics[width=0.47\textwidth]{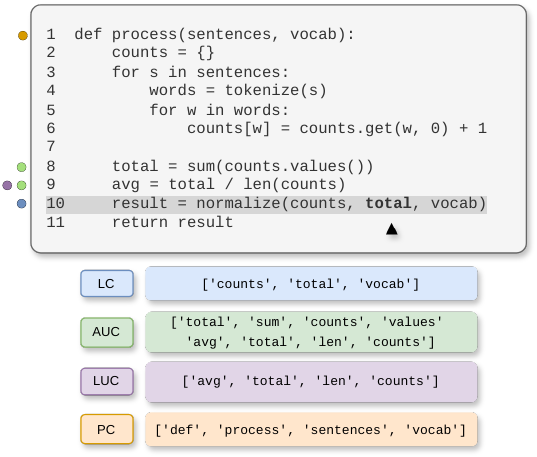}
    \caption{Example of collecting contexts for the argument 'total'.}
    \label{fig:context}
    \vspace{-4mm}
\end{figure}

An example of collecting contexts for the argument 'total' is shown in Figure~\ref{fig:context}. For visual clarity, tokens in the figure are shown without expression-type (ET) encoding, and the local-context (LC) token window is reduced to two tokens (one each side). 

\subsubsection{Embedding Generation}

For each call site with at least two arguments, we extract the four context types (LC, AUC, LUC, and PC) for each argument. Each context instance is treated as a separate sentence composed of its constituent tokens. This process yields four context-specific corpora, one per context type, each containing token sequences derived from argument usages across the corpus.

To better preserve syntactic and semantic roles during tokenization, BugProbe optionally augments tokens with their expression type (ET). An expression-type encoded token appends a lightweight syntactic role (e.g., assignment target, call argument, binary operand) to the original identifier, allowing identical names used in different expression contexts to be distinguished during embedding learning. Table~\ref{tab:expression-tokenization} illustrates common Python expressions and their corresponding token sequences with and without ET encoding.

\begin{table*}[t]
\centering
\caption{Tokenization of common Python expressions with and without expression-type (ET) encoding.}
\label{tab:expression-tokenization}
\begin{tabular}{lll}
\toprule
Expression & non-ET tokens & ET tokens \\
\midrule
\texttt{x} &
\texttt{["x"]} &
\texttt{["x\#Name"]} \\

\texttt{x = y} &
\texttt{["x", "y"]} &
\texttt{["x\#AssignTarget", "y\#Name"]} \\

\texttt{a + b} &
\texttt{["a", "b"]} &
\texttt{["a\#BinOp", "b\#BinOp"]} \\

\texttt{a == b} &
\texttt{["a", "b"]} &
\texttt{["a\#Compare", "b\#Compare"]} \\

\texttt{f(x)} &
\texttt{["f", "x"]} &
\texttt{["f\#CallFunc", "x\#CallArg"]} \\

\texttt{obj.m(x)} &
\texttt{["obj", "m", "x"]} &
\texttt{["obj\#Name", "m\#Attribute", "x\#CallArg"]} \\

\texttt{arr[i]} &
\texttt{["arr", "i"]} &
\texttt{["arr\#Subscript", "i\#Name"]} \\

\texttt{[a, b]} &
\texttt{["a", "b"]} &
\texttt{["a\#List", "b\#List"]} \\
\bottomrule
\end{tabular}
\vspace{-4mm}
\end{table*}
We concatenate these corpora into a single training corpus and train a Word2Vec model using the Continuous Bag-of-Words (CBOW) architecture with a context window size of 5 and an embedding dimensionality of 100. By training on aggregated context signals rather than isolated identifiers, the resulting embeddings capture richer semantic relationships grounded in both lexical usage and structural code context.

\subsubsection{Classifier}

\begin{figure}[t]
    \centering
    \includegraphics[width=\columnwidth]{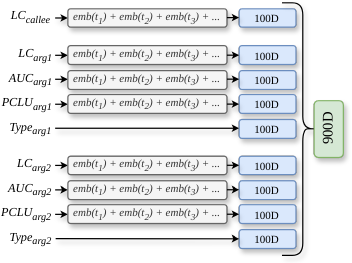}
    \caption{Overview of Vectorization.}
    \label{fig:vectorization}
    \vspace{-2mm}
\end{figure}

Figure~\ref{fig:vectorization} illustrates the feature construction and classification pipeline.
For each training example (positive or negative), we construct a fixed-length representation by encoding
the 9 distinct program fields: one callee context field, two argument contexts (each with three fields),
and two type fields.

Each context field consists of a sequence of tokens. Tokens are embedded using the embedding model,
and the embeddings within a field are aggregated via mean pooling to produce a single 100-dimensional vector.
In contrast, the two type fields are represented using fixed, non-learned vectors rather than the embedding
model.

The resulting nine 100-dimensional vectors---corresponding to the callee context, argument contexts, and type information---are concatenated to form a single 900-dimensional feature vector representing the example.
This vector is then provided as input to a feed-forward multilayer perceptron with a single hidden layer
of size 100. The classifier is then trained to distinguish correct from incorrect argument orderings and is evaluated on a
held-out test set to detect likely argument-swap bugs in previously unseen Python code.
\section{Evaluation}
\label{sec:eval}
This section describes our evaluation procedure and results involving real-world Python code. Our study focuses on the following research questions:

\begin{itemize}
    \item \textbf{\textit{RQ1:}} How effective is our approach in detecting argument-swap bugs?
    \item \textbf{\textit{RQ2:}} Can a simpler name-based check be effective in detecting incorrectly ordered arguments?
    \item \textbf{\textit{RQ3:}} What is the impact of different sources of information? 
    \item \textbf{\textit{RQ4:}} Can we use pre-trained word embeddings in detecting argument-related bugs?
    \item \textbf{\textit{RQ5:}} How does BugProbe perform on real-world bugs?
\end{itemize}
The dataset and replication source are available at
\url{https://zenodo.org/records/21942055}
to support future reproductions.

\subsection{Experimental Setup}
Our dataset comprises 132,739 Python source files collected from the top-1,000 starred GitHub repositories, as described in Section~\ref{sec:data}, parsed using Python's AST module\footnote{\url{https://docs.python.org/3/library/ast.html}} and totaling 34,357,043 lines of code.

Among all qualifying call sites, 1,685,622 have more than one argument and are therefore retained for our evaluation. Each call site contributes one positive and one negative example, yielding 3,371,244 total examples. Of these, 2,696,686 are used for training and 674,558 for testing across the five folds.

To evaluate the baselines, we utilize a 5-fold cross-validation strategy at the \emph{file level}. Specifically, the dataset is randomly partitioned into five disjoint subsets of approximately equal size. During each fold, one subset (20\% of the files) is used as the test set, while the remaining four subsets (80\% of the files) are used for training. This process is repeated five times, such that each subset serves as the test set exactly once. All reported results are averaged over the five folds. For each fold, the Word2Vec embeddings are trained exclusively on that fold's training files, ensuring no information from the test files influences the learned representations. Files containing real-world bugs from our curated benchmark are excluded from this split entirely, and therefore from Word2Vec training as well.

We employ the Gensim implementation of Word2Vec\footnote{\url{https://radimrehurek.com/gensim/models/word2vec.html}} to generate vector embeddings. Classification models are built using machine learning algorithms provided by the Scikit-learn library\footnote{\url{https://scikit-learn.org/stable/}}. All experiments are conducted on a machine equipped with an AMD Ryzen 9 HS series processor and 32~GB of RAM.

\subsection{Evaluation Metrics}
We evaluate detection performance using precision, recall, and the $F_1$ score.

Precision measures the proportion of correct predictions among all predictions made by the classifier, while recall measures the proportion of correctly identified instances among all true instances in the test set. The $F_1$ score, defined as the harmonic mean of precision and recall, is computed as follows:
\[
F_1 = \frac{2 \times \text{precision} \times \text{recall}}{\text{precision} + \text{recall}}
\]

Since the classifier is expected to identify both incorrect and correct code instances, we report precision, recall, and $F_1$ score separately for each class.

\subsection{RQ1: Effectiveness of BugProbe}

\subsubsection{Motivation}
For its best results, DeepBugs benefits from mapping call sites to their corresponding function definitions, as this provides access to formal parameter names that serve as key features. Such mapping is frequently infeasible when definitions reside in different files or external libraries, a challenge that is especially pronounced in dynamically typed languages. In our study of Python code, we found that 89\% of call sites have their definitions located in different files, frequently within third-party libraries. This motivates our research question: can effective bug detection be achieved without any reliance on call--definition mappings, thereby improving robustness and coverage in dynamic language settings?

\subsubsection{Approach}
Both models are trained and evaluated on the same 5-fold cross-validation splits across all call sites. To isolate the effect of call--definition mapping on detection performance, we partition the test results into two subsets: (i) \emph{mapped call sites}, for which a function definition is resolvable within the same file, and (ii) \emph{all call sites}, which includes all call sites regardless of whether a definition is available. Reporting results for each partition separately allows us to compare the two approaches on DeepBugs' most favorable condition --- where it has access to formal parameter name features --- as well as across the complete dataset. Since BugProbe requires no call--definition mapping, its features are identical across both partitions; the distinction affects only how much information is available to DeepBugs.

\subsubsection{Results}
Tables~\ref{tab:mmc} and~\ref{tab:amc} report the performance of DeepBugs and BugProbe on mapped call sites and on the complete dataset, respectively. On mapped call sites, BugProbe consistently outperforms DeepBugs, achieving an accuracy of 87.8\% compared to 84.6\%, corresponding to an absolute improvement of 3.2 percentage points (3.8\% relative), alongside higher per-class precision, recall, F1-score, and improved AUC. The performance gap becomes more pronounced when considering all call sites: BugProbe attains an accuracy of 91.1\%, compared to 81.8\% for DeepBugs, yielding an absolute improvement of 9.3 percentage points (approximately 11.4\% relative). Similar gains are observed across precision, recall, and F1-score, with BugProbe improving F1-scores by approximately 9 percentage points across classes, and increasing AUC from 92.0 to 97.5. When isolating only the unmapped call sites, the gap widens further, with BugProbe achieving an accuracy improvement of 10.2 percentage points over DeepBugs. These results demonstrate that BugProbe not only outperforms DeepBugs when call--definition mappings are available, but also remains robust and highly effective in their absence.

\subsection{RQ2: Evaluating a Name-based Similarity Matching Solution}

\subsubsection{Motivation}
\begin{figure}[t]
    \centering
\includegraphics[width=0.5\textwidth]{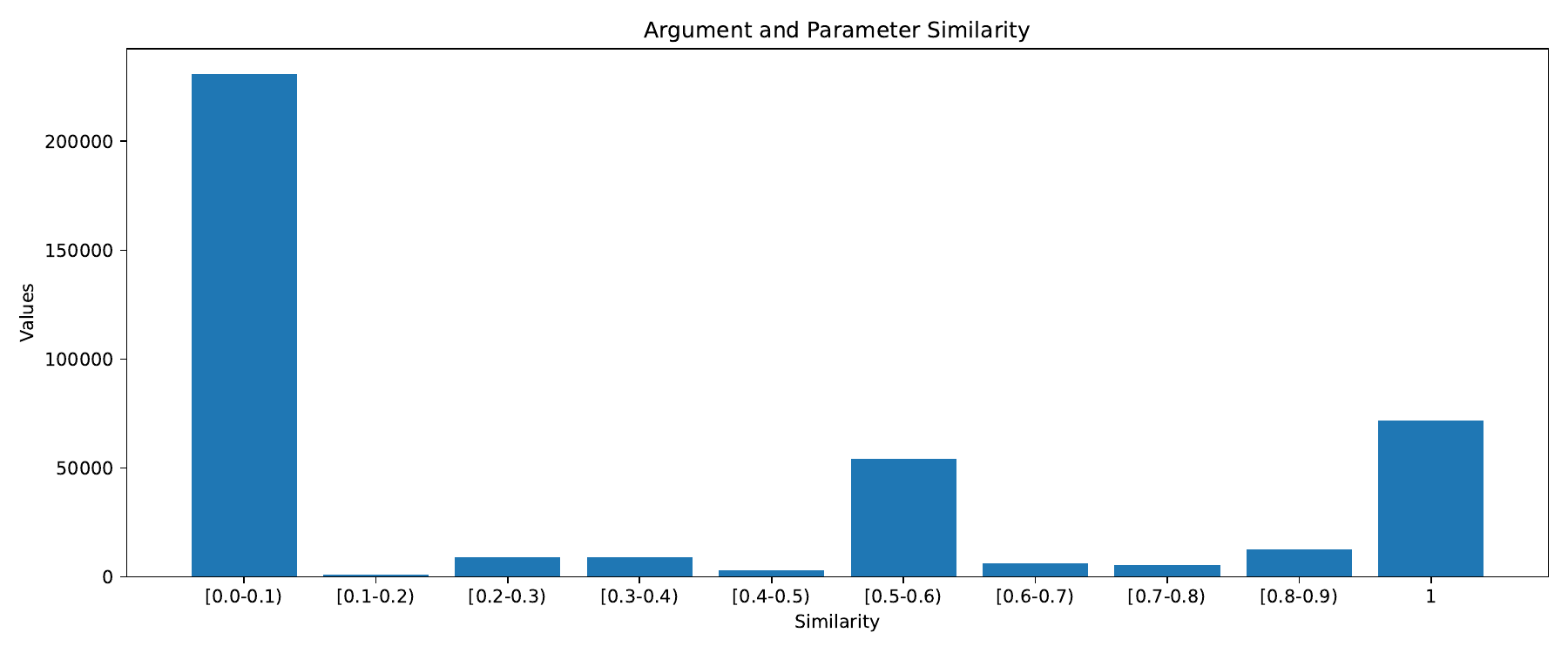}
\vspace{-2mm}
    \caption{Argument–Parameter Lexical Similarity.}
    \label{fig:fisgim1}
    \vspace{-3mm}
\end{figure}

\begin{figure}[t]
    \centering
    \includegraphics[width=0.5\textwidth]{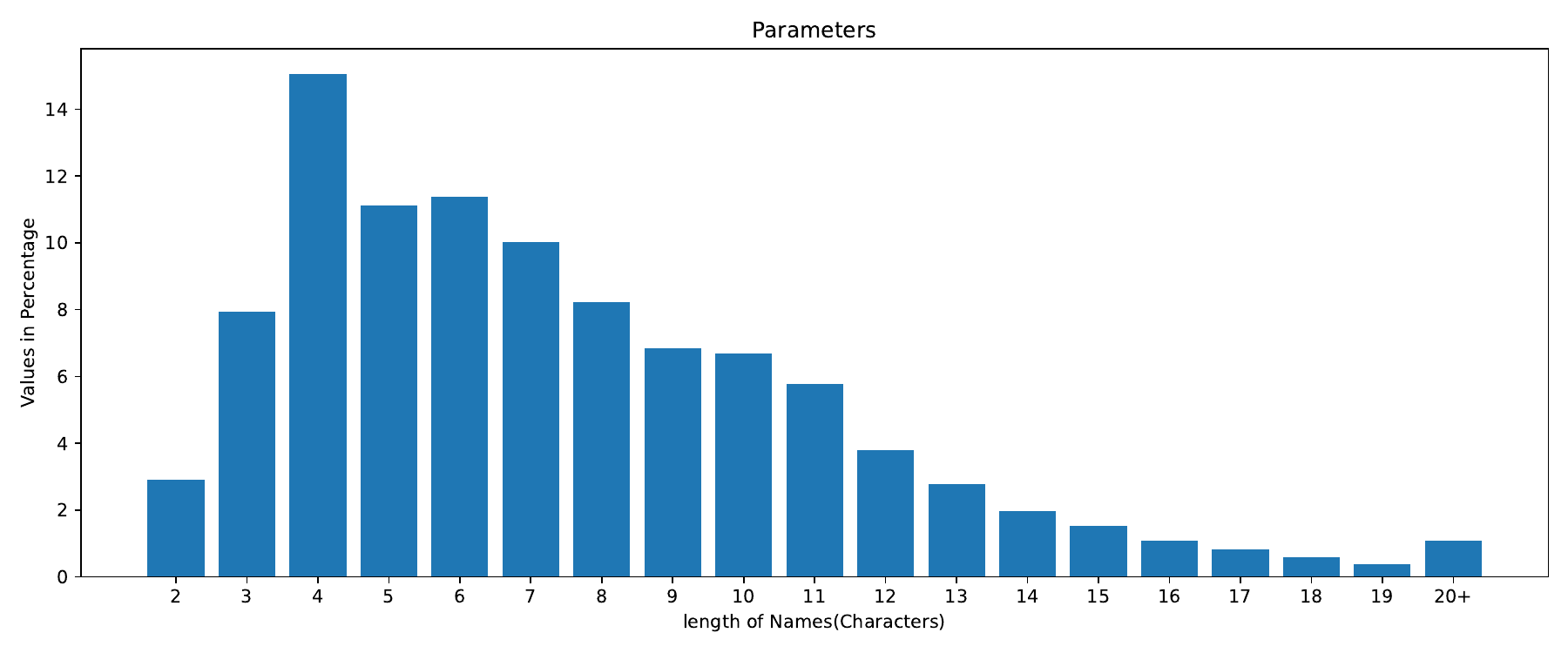}
    \vspace{-2mm}
    \caption{Distribution of Parameter Name Lengths.}
    \label{fig:fisgim2}
    \vspace{-3mm}
\end{figure}

Several prior studies have shown that developers often choose argument names that are lexically similar to their corresponding formal parameter names, and that this similarity can be exploited to detect bugs caused by incorrectly ordered arguments. Figure~\ref{fig:fisgim1} illustrates the distribution of argument--parameter lexical similarity scores across our dataset, and Figure~\ref{fig:fisgim2} shows the distribution of parameter name lengths. Before relying on more complex learning-based techniques, it is important to assess how effective a simple name-based similarity check can be in practice. This research question investigates whether lexical similarity between argument names and parameter names alone is sufficient to reliably detect argument-swap bugs in Python.

\subsubsection{Approach}
We implement a lightweight name-based similarity matching technique that relies solely on lexical overlap between argument names at call sites and their corresponding formal parameter names. Identifier names are first normalized by splitting them into terms using camel-case boundaries, underscores, and non-alphanumeric delimiters.

The lexical similarity between an argument name $arg$ and a parameter name $par$ is computed using a term-based similarity measure adapted from prior work:
{\small
\[
Sim(arg, par) =
\frac{|comterms(arg, par)| + |comterms(par, arg)|}
{|terms(arg)| + |terms(par)|}
\]
}
where $terms(x)$ denotes the set of extracted terms from an identifier name, and $comterms(x,y)$ denotes the set of common terms between two identifiers.

For a call site with two arguments, we compute the average similarity score between each argument and its corresponding parameter. If the resulting score exceeds a fixed threshold (0.5 in our experiments), the call is classified as correctly ordered; otherwise, it is classified as incorrectly ordered. As with other parts of our evaluation, we assume that most code is correct and generate negative examples by synthetically swapping the first two arguments. The implementation follows this procedure directly and does not rely on any learned representations or contextual information.

\subsubsection{Results}
The name-based similarity matching approach achieves high recall but moderate precision. For the incorrect class, it achieves 75.3\% precision and 94.5\% recall (F1 = 83.8\%), effectively flagging most argument-swap bugs but with a notable false-positive rate. For correctly ordered arguments, precision is high (92.6\%) but recall drops to 69.0\% (F1 = 79.1\%), indicating that the check is conservative yet still frequently misclassifies correct calls as incorrect.

Overall, lexical similarity alone yields promising recall but falls short as a reliable standalone detector, and it fundamentally depends on the availability of formal parameter names, which are not always statically resolvable in Python. These limitations motivate the richer, learning-based approaches explored in subsequent research questions.

\begin{table}[htb]
\centering
\caption{Performance Comparison between DeepBugs and BugProbe on Mapped Call Sites (12.7\% of All Call Sites)}
\label{tab:mmc}
\begin{tabular}{lccccc}
\toprule
& \multicolumn{2}{c}{\textbf{DeepBugs}} & \multicolumn{2}{c}{\textbf{BugProbe}} \\
\cmidrule(lr){2-3} \cmidrule(lr){4-5}
\textbf{Metric} & \textbf{Incorrect} & \textbf{Correct} & \textbf{Incorrect} & \textbf{Correct} \\
\midrule
Accuracy  & \multicolumn{2}{c}{84.6} & \multicolumn{2}{c}{87.8} \\
Precision & 85.6 & 83.7 & 86.8 & 88.8 \\
Recall    & 83.3 & 85.9 & 89.1 & 86.5 \\
F1-score  & 84.4 & 84.8 & 87.9 & 87.6 \\
AUC       & \multicolumn{2}{c}{93.7} & \multicolumn{2}{c}{95.6} \\
\bottomrule
\end{tabular}
\vspace{-5mm}
\end{table}

\begin{table}[htb]
\centering
\caption{Performance Comparison between DeepBugs and BugProbe on All Call Sites}
\label{tab:amc}
\begin{tabular}{lccccc}
\toprule
& \multicolumn{2}{c}{\textbf{DeepBugs}} & \multicolumn{2}{c}{\textbf{BugProbe}} \\
\cmidrule(lr){2-3} \cmidrule(lr){4-5}
\textbf{Metric} & \textbf{Incorrect} & \textbf{Correct} & \textbf{Incorrect} & \textbf{Correct} \\
\midrule
Accuracy  & \multicolumn{2}{c}{81.8} & \multicolumn{2}{c}{91.1} \\
Precision & 81.5 & 82.2 & 89.9 & 92.3 \\
Recall    & 82.3 & 81.2 & 92.5 & 89.6 \\
F1-score  & 81.9 & 81.7 & 91.2 & 90.9 \\
AUC       & \multicolumn{2}{c}{92.0} & \multicolumn{2}{c}{97.5} \\
\bottomrule
\end{tabular}
\end{table}


\subsection{RQ3: Impact of Different Sources of Information}

\subsubsection{Motivation}
BugProbe represents call sites and their arguments using multiple complementary sources of contextual information, each capturing a different aspect of how arguments are defined, manipulated, and consumed. Local lexical context provides immediate syntactic cues but is often insufficient for subtle errors such as argument swaps; argument usage context models how values flow through assignments prior to a call; parent and latest usage contexts encode structural information from enclosing constructs; and lightweight type information differentiates literals and primitives, particularly when swapped arguments share the same static type.

Although each source is individually informative, their relative contributions remain unclear. BugProbe can also incorporate expression-typed (ET) token encoding to distinguish identifiers by syntactic role, at the cost of added complexity. This motivates an empirical investigation into the individual and combined impact of these feature groups, and whether ET encoding yields benefits beyond standard tokenization.

\subsubsection{Approach}
To assess the contribution of different sources of information, we conduct an ablation study using five models (M1–M5), each corresponding to a distinct combination of feature groups, as shown in Table~\ref{tab:feature-ablation}. Model M5 corresponds to the full BugProbe configuration, incorporating all available context types: callee local context, argument local context, argument usage context (AUC), parent and latest usage context (PCLU), and argument type information. The remaining models selectively remove one or more feature groups to isolate their effects.

For each model, we evaluate performance with and without expression-typed (ET) token encoding while keeping the training and evaluation protocol identical to previous experiments. This allows us to directly compare the influence of both contextual feature groups and ET encoding on accuracy, F1-score, and AUC.

\begin{table*}[t]
\centering
\caption{Effects of different feature groups with and without expression-typed (ET) token encoding. Each model (M1–M5) corresponds to a distinct combination of feature groups.}
\small
\setlength{\tabcolsep}{4pt}
\renewcommand{\arraystretch}{1.25}

\begin{tabular}{c c c c c c ccc ccc}
\toprule
\multirow{2}{*}{Model} &
\multicolumn{5}{c}{Feature Groups} &
\multicolumn{3}{c}{No ET} &
\multicolumn{3}{c}{ET} \\
\cmidrule(lr){2-6}
\cmidrule(lr){7-9}
\cmidrule(lr){10-12}
& $\mathrm{LC}_{\text{callee}}$
& $\mathrm{LC}_{\text{args}}$
& $\mathrm{AUC}_{\text{args}}$
& $\mathrm{PCLU}_{\text{args}}$
& $\mathrm{Type}_{\text{args}}$
& Acc & F1 & AUC
& Acc & F1 & AUC \\
\midrule

M1 &  & \fdot &  &  &  
   & 62.06 & 62.07 & 68.46 & 67.11 & 67.86 & 75.28 \\

M2 & & & \fdot & & \fdot
   & 65.61 & 64.10 & 73.95 & 65.77 & 66.66 & 73.96 \\

M3 & \fdot & \fdot &  &  &  
   & 89.85 & 89.98 & 97.08 & 90.63 & 90.85 & 97.52 \\

M4 & \fdot & \fdot & \fdot & \fdot &  
   & 90.54 & 90.61 & 97.23 & 90.98 & 91.13 & 97.52 \\

M5 & \fdot & \fdot & \fdot & \fdot & \fdot
   & 90.68 & 90.71 & 97.26 & 91.08 & 91.21 & 97.51 \\
\bottomrule
\end{tabular}
\label{tab:feature-ablation}
\vspace{-3mm}
\end{table*}
\subsubsection{Results}
Table~\ref{tab:feature-ablation} shows that using a single source of information is insufficient for accurate argument-swap detection. Models M1 and M2, which rely only on local argument context or argument usage context (with type) respectively, achieve F1-scores of 62.07\% and 64.10\% without ET encoding, with AUC values below 75\%. Compared to the full BugProbe model (M5), this corresponds to a decrease of approximately 26--28 percentage points in F1-score, indicating that isolated context signals fail to capture the semantic relationships necessary for identifying incorrect argument orderings.

Combining callee-level local context with argument local context (M3) leads to a large performance gain. M3 achieves an F1-score of 89.98\% and an AUC of 97.08\% without ET encoding, improving F1 by approximately 27.9 percentage points over M1. Adding longer-range contextual information through argument usage context and parent/latest usage context (M4) yields an F1-score of 90.61\% and an AUC of 97.23\%, which is within 0.1 percentage points of the full model. The complete BugProbe configuration (M5) achieves the highest overall performance, with an F1-score of 90.71\% and an AUC of 97.26\%, confirming that each additional context source contributes incremental but measurable improvements.

Expression-typed (ET) token encoding provides consistent but modest gains across most configurations. The largest absolute improvement is observed for M1 (from 62.07\% to 67.86\% F1, a gain of 5.79 percentage points), while gains for context-rich models such as M4 and M5 are smaller (approximately 0.5 percentage points). These results suggest that while ET encoding does not fundamentally change model behavior, it provides useful syntactic disambiguation that complements contextual information and slightly improves detection accuracy.

\subsection{RQ4: Performance Comparison with Pre-trained Word Embedding Models}
\subsubsection{Motivation}
BugProbe relies on Word2Vec embeddings trained on the target corpus, capturing project-specific naming conventions. While pre-trained code embeddings have shown strong performance on many software engineering tasks, it remains unclear whether their broader representational capacity benefits argument-ordering bug detection specifically.

\subsubsection{Approach}
We replace BugProbe's Word2Vec component with representations from CodeBERT~\cite{feng2020codebert}, a transformer-based model built on BERT~\cite{devlin2019bert} and pre-trained on natural language and source code, keeping all other aspects of the framework unchanged. We refer to this variant as $BugProbe_{\text{CodeBERT}}$.

\subsubsection{Results}
$BugProbe_{\text{CodeBERT}}$ achieves an accuracy of 81.0\%, an $F_1$ score of 81.5\% for incorrect examples (82.6\% precision, 78.7\% recall) and 80.6\% for correct ones (79.6\% precision, 83.4\% recall), and a ROC-AUC of 89.6\%. These results are competitive but fall short of the original BugProbe across all metrics, at greater computational cost, suggesting that BugProbe's lightweight Word2Vec embeddings better capture the fine-grained, usage-specific signals this task requires~\cite{ding2022pretrained}.

\subsection{RQ5: How Does BugProbe Perform on Real-World Bugs?}
\subsubsection{Motivation}
The synthetic evaluation in RQ1 relies on negative examples generated by swapping argument pairs. Although this enables large-scale evaluation and follows prior work, it does not directly show whether a technique can detect bugs that occur in practice. Real-world argument-swap bugs may differ from synthetic ones in subtle ways. Therefore, we evaluate BugProbe on our benchmark of 55 real-world bugs to assess its practical utility.
\subsubsection{Approach}
We use the proposed benchmark of 55 argument-swap bugs. Each bug corresponds to a call site where the arguments were manually confirmed to be incorrectly ordered. For each case, we classify the original buggy call as a negative example and the corrected version from the fixing commit as a paired positive example. Both versions are embedded and classified using models trained only on the synthetic dataset, without fine-tuning on the real-world benchmark. A bug is considered detected if the classifier assigns a higher likelihood of incorrectness to the buggy version than to the corrected one. We apply the same procedure to both BugProbe and DeepBugs using the model checkpoints from the synthetic evaluation.
\subsubsection{Results}
Table~\ref{tab:rq5} reports the performance of DeepBugs and BugProbe on the real-world benchmark. Both approaches perform lower than in the synthetic evaluation, reflecting the greater diversity and subtlety of real-world bugs. Nevertheless, BugProbe consistently outperforms DeepBugs, achieving 80.9\% accuracy compared to 53.6\%, and improving the $F_1$ score by approximately 20 percentage points for the incorrect class (79.6\% vs.\ 59.2\%) and approximately 36 percentage points for the correct class (82.1\% vs.\ 46.3\%). BugProbe also achieves a substantially higher AUC, 90.8 compared to 55.1. These results suggest that BugProbe generalizes better to real argument-swap bugs and provides stronger detection capability than the baseline.

\begin{table}[htb]
\centering
\caption{Performance Comparison between DeepBugs and BugProbe on Real-World Bugs (55 bugs)}
\label{tab:rq5}
\begin{tabular}{lccccc}
\toprule
& \multicolumn{2}{c}{\textbf{DeepBugs}} & \multicolumn{2}{c}{\textbf{BugProbe}} \\
\cmidrule(lr){2-3} \cmidrule(lr){4-5}
\textbf{Metric} & \textbf{Incorrect} & \textbf{Correct} & \textbf{Incorrect} & \textbf{Correct} \\
\midrule
Accuracy  & \multicolumn{2}{c}{53.6} & \multicolumn{2}{c}{80.9} \\
Precision & 52.9 & 55.0 & 85.4 & 77.4 \\
Recall    & 67.3 & 40.0 & 74.6 & 87.3 \\
F1-score  & 59.2 & 46.3 & 79.6 & 82.1 \\
AUC       & \multicolumn{2}{c}{55.1} & \multicolumn{2}{c}{90.8} \\
\bottomrule
\end{tabular}
\vspace{-3mm}
\end{table}

\subsection{Error Analysis}
\label{sec:error-analysis}

To better understand BugProbe's remaining limitations, we qualitatively analyzed its false positives and false negatives.

False positives commonly arise when arguments are semantically related or frequently co-occur even when correctly ordered, such as \texttt{width}/\texttt{height} or \texttt{x}/\texttt{y}. Swapping such arguments may not substantially alter the surrounding context signals, making correct-vs-incorrect orderings hard to distinguish confidently; this mirrors the high-recall, low-precision behavior of the name-based baseline in RQ2.

False negatives typically occur when argument names are overly generic (e.g., \texttt{a}, \texttt{b}, \texttt{val}, \texttt{data}) or derived from similar expressions, leaving neither lexical similarity nor contextual usage with sufficient discriminative signal.
\section{Discussion}
\label{sec:threats}

\subsection{Efficiency and Practical Applicability}

BugProbe's pipeline consists of several stages: parsing, context collection, vector generation, training, and testing. Parsing requires 15~ms on average, and context collection adds a further 1.1~ms per example.

Model training requires 43 minutes on our machine when considering all call sites in the dataset. In contrast, testing is considerably faster, requiring only 4 minutes in total for 674,558 examples. Overall, BugProbe can determine whether the arguments of a call site are correctly ordered or generate a warning for incorrectly ordered arguments within 1.1 ms on average per instance. These results suggest that BugProbe is efficient enough to be integrated into development environments and used in near real-time, for example, to provide feedback as developers write code.

\subsection{Threats to Validity}

\textbf{Internal validity.} Our primary evaluation relies on synthetic negative examples generated by swapping the first two arguments of call sites. While widely adopted in prior work, not all synthetic swaps necessarily correspond to real bugs, and some swapped calls may remain semantically correct. Since the same generation procedure is applied to all evaluated approaches, this threat does not bias the comparative results. We further evaluate BugProbe on a curated benchmark of 55 real-world argument-swap bugs, which directly assesses detection on organically occurring defects and substantially mitigates the above-mentioned concern, though at 55 bugs it remains small relative to the synthetic set and results should be read as indicative rather than statistically comprehensive.

\textbf{Construct validity.} We frame argument-ordering bug detection as binary classification and evaluate performance using precision, recall, and $F_1$ score. These metrics capture detection effectiveness but not developer-perceived usefulness or bug severity; user studies could offer further insight into practical impact.

\textbf{External validity.} Our experiments use a large corpus of open-source Python projects from GitHub. While the dataset spans diverse domains, results may not generalize to proprietary codebases or other languages, particularly those with stronger static typing or different naming conventions. Python nonetheless remains a challenging and widely used dynamically typed language, making it a meaningful target for this study.

\textbf{Conclusion validity.} BugProbe relies on Word2Vec embeddings, chosen to ensure a fair comparison with DeepBugs; alternative embedding architectures may yield different results. Our findings on pre-trained embeddings are therefore specific to the evaluated models rather than a general limitation of pre-trained representations.
\section{Conclusion}
\label{sec:conclusion}
In this paper, we investigate the problem of detecting incorrectly ordered arguments in Python programs and propose BugProbe, a machine learning–based approach for identifying argument-ordering bugs. BugProbe frames the task as binary classification and uses learned representations derived from multiple contextual signals, including local lexical context, argument usage patterns, structural context, and lightweight type information, capturing both local and long-range dependencies relevant to argument ordering. Evaluations on synthetic and real-world bug datasets show that BugProbe consistently outperforms DeepBugs and other baseline techniques. Crucially, BugProbe requires no explicit mappings between call sites and their definitions, making it well suited to dynamically typed languages where such mappings are often unavailable. An ablation study further shows that combining multiple contextual feature groups is key to strong performance, while expression-typed token encoding yields modest but consistent gains.

In future work, we plan to explore richer type information from static type inference tools~\cite{10.1145/3510003.3510038, aman2026typify}, extend BugProbe to other semantic bug patterns, and compare its effectiveness against large language models (LLMs) and additional baselines such as SWAPD~\cite{scott2020out} and NALIN~\cite{patra2022nalin} for source-code bug detection.

\bibliographystyle{IEEEtran}
\bibliography{refs}
\end{document}